\documentclass{article}
\usepackage{jheppub}

\usepackage{ifpdf}

\usepackage{afterpage}

\usepackage{amssymb}
\usepackage{amsmath}
\usepackage{graphicx}
\usepackage{longtable}
\usepackage{verbatim}
\usepackage{amsfonts}
\usepackage{color, colortbl}
\definecolor{LightCyan}{rgb}{0.88,1,1}
\definecolor{LightViolet}{rgb}{1, 0.88,1}
\definecolor{LightGreen}{rgb}{0.88, 1, 0.88}

\usepackage{color, colortbl}
\definecolor{LightCyan}{rgb}{0.88,1,1}
\definecolor{LightViolet}{rgb}{1, 0.88,1}
\definecolor{LightGreen}{rgb}{0.88, 1, 0.88}

\newcommand{\bear}{\begin{array}}
\newcommand{\ear}{\end{array}}

\newcommand{\beq}{\begin{eqnarray}}
\newcommand{\eeq}{\end{eqnarray}}
\newcommand{\beqa}{\begin{eqnarray}}
\newcommand{\eeqa}{\end{eqnarray}}
\newcommand{\no}{\nonumber}

\def\OMIT#1{{}}
\newcommand{\lsim}{\mathrel{\rlap{\lower4pt\hbox{\hskip1pt$\sim$}}
    \raise1pt\hbox{$<$}}}         
\newcommand{\gsim}{\mathrel{\rlap{\lower4pt\hbox{\hskip1pt$\sim$}}
    \raise1pt\hbox{$>$}}}         

\newcommand{\be}{\begin{equation}}
\newcommand{\ee}{\end{equation}}
\newcommand{\ba}{\begin{eqnarray}}
\newcommand{\ea}{\end{eqnarray}}

\def\lsim{\mathrel{\rlap{\lower4pt\hbox{\hskip1pt$\sim$}}
    \raise1pt\hbox{$<$}}}         
\def\gsim{\mathrel{\rlap{\lower4pt\hbox{\hskip1pt$\sim$}}
    \raise1pt\hbox{$>$}}}         

\begin{document}

\vspace*{-30mm}

\title{\boldmath 2:1 for Naturalness at the LHC?}

\author[a]{Nima Arkani-Hamed,}
\author[a]{Kfir Blum,}
\author[b,c]{Raffaele Tito D'Agnolo,}
\author[d]{JiJi Fan}
\affiliation[a]{School of Natural Sciences, Institute for Advanced Study, Princeton, NJ 08540, USA}
\affiliation[b]{Scuola Normale Superiore and INFN, Piazza dei Cavalieri 7, 56126, Pisa, Italy}
\affiliation[c]{CERN, European Organization for Nuclear Research, Geneva, Switzerland}
\affiliation[d]{Department of Physics, Princeton University, Princeton, NJ 08540, USA}

\vspace*{1cm}

\abstract{A large enhancement of a factor of 1.5 - 2 in Higgs production and decay in
the diphoton channel, with little deviation in the $ZZ$ channel, can only
plausibly arise from a loop of new charged particles with large couplings
to the Higgs. We show that, allowing only new fermions with marginal
interactions at the weak scale, the required Yukawa couplings for a factor
of 2 enhancement are so large that  the Higgs quartic coupling is pushed to
large negative values in the UV, triggering an unacceptable vacuum
instability far beneath the 10 TeV scale. An enhancement by a factor
of 1.5 can be accommodated if the charged particles are lighter than 150
GeV, within reach of discovery in almost all cases in the 8 TeV run at the
LHC, and in even the most difficult cases at 14 TeV.
Thus if the diphoton enhancement survives further scrutiny, and no charged
particles beneath 150 GeV are found, there must be new bosons far beneath
the 10 TeV scale. This would unambiguously rule out a large class of fine-tuned
theories for physics beyond the Standard Model, including split SUSY and
many of its variants, and provide strong circumstantial evidence
for a natural theory of electroweak symmetry breaking at the TeV scale.
Alternately, theories with only a single fine-tuned Higgs and new fermions
at the weak scale, with no additional scalars or gauge bosons up to a
cutoff much larger than the 10 TeV scale, unambiguously predict that the
hints for a large diphoton enhancement in the current data will disappear.}
\maketitle

\section{Introduction}
The recent announcement of the discovery of the Higgs particle by ATLAS and
CMS represents a triumphant milestone for fundamental physics~\cite{ATLAS,
CMS}. All eyes are now turned to examining the properties of the Higgs in
detail, looking for possible deviations from Standard Model (SM) behavior.
Indeed, in these early days, both ATLAS and CMS have an accumulating hint
of an anomaly. 
While $\sigma \times BR (h \to Z Z^*)$ and $\sigma \times BR (h \to W W^*)$ seem compatible with the SM\footnote{The latter is admittedly an experimentally difficult channel. Note also that while CMS results hint to some deficit in $h\to VV$, ATLAS shows a potential excess.}, there appears to be a significant enhancement in the diphoton channel $\sigma \times BR(h \to \gamma \gamma)$, that may be as high as a factor of 2 above the SM expectation: 
%
%
\ba
\mu_{\gamma \gamma} &=& \frac{\sigma \times BR(h \to \gamma \gamma)}{\sigma
\times BR(h \to \gamma \gamma)_{SM}} \sim 1.5 - 2,\label{eq:gamexp}\\
\mu_{VV} &=& \frac{\sigma \times BR(h \to VV)}{\sigma
\times BR(h \to VV)_{SM}} \sim 1.\label{eq:vvexp}
\ea
Of course the most conservative and likely possibility is that this modest
excess will not survive further scrutiny, and will diminish when all the
2012 data is analyzed. It is nonetheless interesting to contemplate the
sorts of new physics that could be responsible for such a large deviation
in $\sigma \times BR(h \to \gamma \gamma)$ while leaving $\sigma \times Br
(h \to ZZ^*, W W^*)$ essentially unaltered.

While it is possible, in principle, to satisfy
Eqs.~(\ref{eq:gamexp}-\ref{eq:vvexp}) by only adjusting the tree-level
couplings of the Higgs to SM particles, we find this possibility rather
unlikely for the following simple reason. Assuming that the only
modification is via the SM tree-level couplings, then for $m_h=125$~GeV we
have $\mu_{\gamma\gamma}^{(\rm
tree)}\approx\left(1.28-0.28\frac{r_t}{r_V}\right)^2\times\mu_{VV}^{(\rm
tree)}$, where $r_t,r_V$ are the ratio of the couplings of the higgs to the
top and the $W/Z$ relative to the SM couplings. Now in order to obtain, for
instance, $\mu_{\gamma\gamma}=1.5~\mu_{VV}$, there are two solutions: i.
$(r_t/r_V)\approx0.2$
, or ii. $(r_t/r_V)\approx9$.\footnote{It is worth recalling that $r_V>1$
can only be realized in models with doubly-charged
scalars~\cite{Low:2009di}.}
Both of these solutions are highly implausible: allowing an order of
magnitude modification to the couplings, it is unlikely that the
$\sim$125~GeV resonance found at the LHC should have production and decay
rates that are all-in-all broadly consistent with the SM Higgs boson.

We conclude that Eqs.~(\ref{eq:gamexp}-\ref{eq:vvexp}) most likely require
a loop contribution from new particles, enhancing $h \to \gamma\gamma$.
Indeed, a large number of groups have explored this possibility, with
additional scalars, vector-like fermions and gauge bosons of various
types~\cite{Low:2009di, Carena:2012xa}. Our purpose in this note is not to
rehash these arguments, but to point out that such a large $\mu_{\gamma
\gamma}$ has a profound implication for the deepest question that confronts
us at the TeV scale: {{\textbf{Is electroweak symmetry breaking natural?}}

Natural theories of electroweak symmetry breaking are expected to have a
plethora of new particles at the weak scale, associated with a solution to
the hierarchy problem. Some of these particles could be responsible for the
observed diphoton enhancement, though this does not automatically occur in
the most minimal models. For instance among the minimal supersymmetric SM
(MSSM) superparticles, a non-negligible effect can only naturally arise
from very light charginos, but even pushing the relevant parameters to
their limits one finds $\mu_{\gamma\gamma}\lsim1.25$ and, more typically,
$\mu_{\gamma\gamma}<1.1$~\cite{Blum:2012ii}. Combining tree-level Higgs
mixing with loop corrections from charginos and stops can boost
$\mu_{\gamma\gamma}>1.5$, but still keeps
$(\mu_{\gamma\gamma}/\mu_{VV})\lesssim1.4$~\cite{Blum:2012ii}. Other possibilities, like e.g.
light staus with extreme left-right mixing, can be
realized~\cite{Carena:2011aa, Carena:2012gp} but come at the cost of
fine-tuning.

As is well-known, the concept of naturalness has been under some pressure from
a variety of directions, and in the past decade new possibilities
for physics beyond the SM have been explored. The idea is that
the Higgs is fine-tuned to be light, 
as a less-dramatic counterpart to
Weinberg's anthropic explanation of the smallness of the cosmological
constant~\cite{Weinberg:1987dv}. Once naturalness is abandoned, there seems to be no need for any
new physics at all at the weak scale. However, aside from naturalness
itself, this seems to throw out the successes of the best natural theories
we have, with low-energy supersymmetry: the beautifully precise prediction
of gauge-coupling unification, and WIMP dark matter. It was therefore
interesting to find that these successes could easily be preserved in ``split" SUSY~\cite{ArkaniHamed:2004fb,Giudice:2004tc,ArkaniHamed:2004yi,Wells:2004di}, where all the scalars of SUSY are taken to be heavy but
the fermions are light, protected by a chiral symmetry.

Split SUSY is a simple example of a class of fine-tuned
theories for physics beyond the SM. These models tend to be
more constrained and predictive in their structure than many natural
theories.
The main reason is that  arbitrary fine-tunings are not allowed; any
fine-tuning should have a clear ``environmental" purpose.  If we consider
a completely generic theory with many interacting scalars, fermions and
gauge fields, a separate fine-tuning is needed for every light scalar. But
additional scalars beyond the Higgs serve
no ``environmental" purpose. Thus in this framework we don't expect any new
light scalars beyond the Higgs. Additional gauge fields would have to be
higgsed by fine-tuned scalars\footnote{We do not consider the baroque possibility that additional gauge groups are broken by
technicolor-like interactions while the SM gauge symmetry is broken by a
fine-tuned Higgs.}, so we don't expect new gauge bosons either. Thus, this
restricted class of fine-tuned theories can only include new fermions, with
no new scalars or gauge fields, up to some scale $\Lambda_{UV} \gg $ TeV.
As an example, in ``minimally split" SUSY~\cite{ArkaniHamed:2006mb}, we
expect a loop factor splitting between scalars and gauginos. Here the
cut-off of the effective theory $\Lambda_{UV}$ is the mass of the heavy
scalars, with $10$ TeV $ < \Lambda_{UV} < 10^3$ TeV.

Consider the diphoton enhancement in these theories. With the Higgs as the only new scalar, we cannot even entertain the possibility of tree-level modifications giving $r_{V}\neq1$: this route is not only implausible, but impossible. We could, in principle, modify $r_t$ through fermion mixing. However, with $r_t\approx0.2$ there would be no Higgs signal at all, whereas  $r_t\approx9$ would not be perturbative. Thus we can only rely on loop effects from new fermions with Yukawa couplings to the Higgs.

The minimal version of split SUSY cannot give a big enough effect --
indeed, the only source for enhancement is the same chargino loop as in
natural SUSY. Thus a large enhancement of 1.5 - 2 immediately rules out
this version of split SUSY. We can however certainly imagine extra fermions
near the TeV scale; a collection of fermions can have their masses
protected by a common chiral symmetry and set by the same scale.

In what follows we ask whether the recent LHC data can be explained in a
framework of this sort. We show that restricting to un-natural models with
only new fermions immediately leads us to a very narrow set-up with sharp
theoretical and experimental implications: (1)  new, vector-like,
un-colored fermions with electroweak quantum numbers must exist and be very
light, within the range $100-150$~GeV; (2) the cut-off scale of the theory
where additional bosonic degrees of freedom must kick in, cannot be high
and is in fact bounded by $\Lambda_{UV}\lsim1-10$ TeV. The cut-off can be
somewhat increased but only at the expanse of significant model-building
gymnastics, which further destroys any hope of perturbative gauge coupling
unification.

\section{The diphoton rate}
A fermionic loop contribution enhancing the Higgs-diphoton coupling requires vector-like representations and large Yukawa couplings to the Higgs boson. This has important ramifications for the consistency of the theory at high scale. To see this, note that in the presence of a new fermion $f$ with electric charge $Q$, the $h\to\gamma\gamma$ partial width reads\footnote{At leading-log plus leading finite-mass correction; see e.g.~\cite{Carena:2012xa} for a recent discussion.}
\beq\label{eq:diph}\frac{\Gamma(h\to\gamma\gamma)}{\Gamma(h\to\gamma\gamma)_{SM}}\approx\left|1+\frac{1}{\mathcal{A}^\gamma_{SM}}Q^2\,\frac{4}{3}\left(\frac{\partial\log m_f}{\partial \log v}\right)\left(1+\frac{7\,m_h^2}{120\,m_f^2}\right)\right|^2,\eeq
with $\Gamma(h\to\gamma\gamma)_{SM}=\left(\frac{G_F\alpha^2m_h^3}{128\sqrt{2}\pi^3}\right)\left|\mathcal{A}^\gamma_{SM}\right|^2$ and\footnote{At leading-log, the SM amplitude is given by the top quark and $W$ boson contributions to the QED beta function, $\left(\mathcal{A}_{SM}^\gamma\right)_{\rm leading-log}= b_t+b_W=+\left(4/3\right)^2-7$. Finite mass corrections modify this prediction slightly to $\mathcal{A}_{SM}^\gamma=-6.49$.} $\mathcal{A}_{SM}^\gamma=-6.49$.
Constructive interference between the SM and the new fermion amplitude requires electroweak symmetry breaking to contribute negatively to the mass of the new fermion. Thus $f$ must be part of a vector-like representation with an electroweak-conserving source of mass.

The basic building block is then the charged vector-like fermion mass matrix, 
\beq\label{eq:MM}\mathcal{L}_M=-\left(\psi^{+Q}\;\chi^{+Q}\right)\left(\bear{cc}m_\psi&\frac{yv}{\sqrt{2}}\\\frac{y^cv}{\sqrt{2}}&m_\chi\ear\right)\left(\bear{c}\psi^{-Q}\\\chi^{-Q}\ear\right)+cc,\eeq
with the Higgs VEV given by $\langle H\rangle=v/\sqrt{2}=174$~GeV. 
Eq.~(\ref{eq:MM}) contains one physical phase, $\phi=\arg\left(m^*_\psi m^*_\chi yy^c\right)$, that cannot be rotated away by field redefinitions. It is straightforward to show that $\phi=0$ maximizes the effect we are after, making $\phi\neq0$ an un-illuminating complication for our current purpose. Hence for simplicity we assume $\phi=0$ in what follows. We are then allowed to take all of the parameters in Eq.~(\ref{eq:MM}) to be real and positive. The two Dirac mass eigenvalues are split by an amount 
\ba\label{eq:mspl} m_2=m_1\left(1+\sqrt{\Delta_v^2+\Delta_y^2+\Delta_m^2}\right),\;\;\Delta_v^2=\frac{2yy^cv^2}{m_1^2},\;\;\Delta_y^2=\frac{\left(y-y^c\right)^2v^2}{2m_1^2},\;\;\Delta_m^2=\frac{(m_\psi-m_\chi)^2}{m_1^2}.\ea
Using Eq.~(\ref{eq:diph}) and assuming that the diphoton rate enhancement comes from changing the partial width $\Gamma(h\to\gamma\gamma)$, with no change to the gluon fusion production cross section, we have\footnote{In Eq.~(\ref{eq:diphrate}), for clarity, we neglected sub-leading finite-mass terms that amount to $<10\%$ correction for $m_f>100$~GeV. However, we keep these terms in our plots.} 
\ba\label{eq:diphrate}\mu_{\gamma\gamma}=\frac{\Gamma(h\to\gamma\gamma)}{\Gamma(h\to\gamma\gamma)_{SM}}\approx\left|1+0.1\,\mathcal{N}\,Q^2\Delta_v^2\left(1+\sqrt{\Delta_v^2+\Delta_y^2+\Delta_m^2}\right)^{-1}\right|^2,\ea
where we generalized to $\mathcal{N}$ copies of~(\ref{eq:MM}). Noting the  LEPII constraint $m_1\gsim100$~GeV, we immediately see that large Yukawa couplings are required in order to achieve a noticeable effect, at least for common charge assignments $Q^2\leq1$. Even if we maximize the effect by tuning $\Delta_m=\Delta_y=0$ (via $m_\psi=m_\chi,\,y=y^c$), an enhancement of $\mu_{\gamma\gamma}\geq1.5$ still requires $yy^c\geq\left(\frac{0.86}{\mathcal{N}\,Q^2}\frac{m_1}{100~{\rm GeV}}\right)^2$.

Before pursuing further the implications of Eq.~(\ref{eq:diphrate}), we pause to point out that we find it implausible for colored particles  (either fermions or bosons, for that matter) to deliver the effect we are after. 
For colored fermions, the gluon fusion rate is approximately given by an equation similar to (\ref{eq:diph}), but replacing $\left(4N_cQ^2/3\mathcal{A}_{SM}^\gamma\right)\to2t_c$, where $t_c$ and $N_c$ are the color representation constant and dimension. A diphoton width enhancement, $\Gamma(h\to\gamma\gamma)/\Gamma(h\to\gamma\gamma)_{SM}=|1+\delta|^2$, would lead to a digluon effect $\mu_{GG}\approx\left|1-9.7(t_c/N_cQ^2)\delta\right|^2$, going through to the $ZZ,WW$ channels as $\mu_{VV}\approx\mu_{GG}$. For scalars (vector bosons), we would simply rescale $\delta$ by a factor of 4 $\left(-\frac{4}{21}\right)$, arriving at the same result. For example, $Q=2/3$ particles in the {\bf 3} of color would give $\mu_{GG}\approx|1-3.6\,\delta|^2$. To accommodate both of Eqs.~(\ref{eq:gamexp}-\ref{eq:vvexp}) in this case, one would need -- similar to our discussion of tree-level solutions -- to accept large distortions of the SM couplings that conspire to leave a moderately small net observable effect. In Fig.~\ref{fig:colornocolor} we illustrate this point, by plotting $\mu_{\gamma\gamma}$ and $\mu_{GG}$ as a function of the diphoton amplitude modification $\delta$, for $Q=2/3$ particles. For uncolored particles (smooth) we have $\mu_{GG}=\mu_{VV}=1$ and $\mu_{\gamma\gamma}=|1+\delta|^2$, while for particles in the ${\bf 3}$ of color (dashed) we have $\mu_{GG}=\mu_{VV}=|1-3.6~\delta|^2$ and $\mu_{\gamma\gamma}=|1+\delta|^2|1-3.6~\delta|^2$. It is obvious from the plot that substantial tuning is required for the colored solution to roughly satisfy Eqs.~(\ref{eq:gamexp}-\ref{eq:vvexp}). Note that while there are two separate colored solutions to $\mu_{\gamma\gamma}\sim1.5-2$, one with $\delta\approx+0.5$ and one with $\delta\approx-0.15$, the former would imply some  $ZZ$ suppression, $\mu_{VV}\sim0.6-0.8$, while the latter would greatly overshoot the SM value $\mu_{VV}\sim2-3$. We therefore discard the possibility of colored particles for addressing the diphoton rate anomaly, at least for electric charge assignments $Q^2\leq1$.
\begin{figure}[!h]\begin{center}
\includegraphics[width=0.6\textwidth]{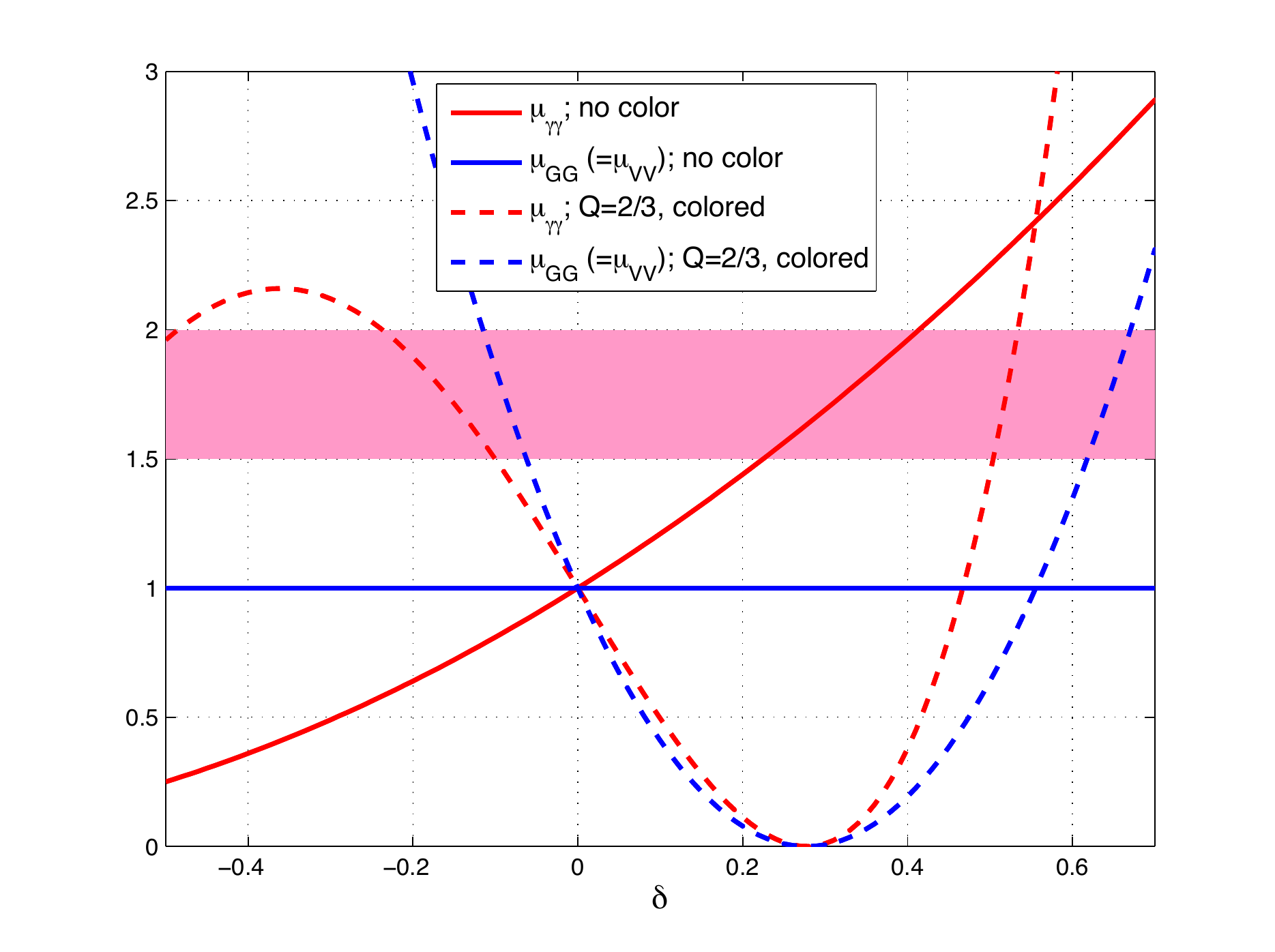}
\end{center}
\caption{Uncolored (smooth lines) and colored (dashed lines) particles, for generating a diphoton partial width enhancement $\Gamma(h\to\gamma\gamma)=\Gamma(h\to\gamma\gamma)_{SM}|1+\delta|^2$.}
\label{fig:colornocolor}
\end{figure}%

We now turn our attention to Eq.~(\ref{eq:diphrate}) and to the large Yukawa couplings that it requires (given some reasonable assumptions about the sorts of multiplets we allow), in order to give an enhancement of $\mu_{\gamma \gamma} \sim 1.5 - 2$. In fact, the needed Yukawa couplings are so large that, unless the new particles are extremely light, the Higgs quartic coupling $\lambda$ is rapidly driven negative at high scales. Importantly, if we assume only fermions up to high scale, the addition of any other fermions only drives the quartic even more negative. Thus vacuum stability becomes an important constraint.  At some scale $\Lambda_{UV}$, $\lambda$ gets so negative that the tunneling rate through false vacuum bubbles of size $\Lambda_{UV}^{-1}$ becomes less than the age of the universe. We define $\Lambda_{UV}$ as the cut-off scale of the (un-natural) theory: here, new bosonic fields must kick in to remedy the instability. 

To substantiate these statements, we next consider two concrete examples. Our Higgs field transforms as $H\sim(1,2)_{\frac{1}{2}}$. It remains to assign $SU(2)$ representations to the fermions in Eq.~(\ref{eq:MM}). 
\paragraph{Vector doublets + singlets (``vector-like lepton"): $\psi,\psi^c\sim(1,2)_{\pm\frac{1}{2}},\;\;\chi,\chi^c\sim(1,1)_{\mp1}.$} 
The Lagrangian leading to (\ref{eq:MM}) is
\ba\label{eq:doubL}-\mathcal{L}=m_\psi\psi\psi^c+m_\chi\chi\chi^c+yH\psi\chi+y^cH^\dag\psi^c\chi^c+cc.\ea
There are two charged Dirac fermions $L_{1,2}$ with masses $m_{L_{1,2}}$ ($m_{L_1} < m_{L_2}$), separated as in Eq.~(\ref{eq:mspl}), and one neutral Dirac fermion $N$ with mass $m_{\psi}$. Given $m_{L_{1,2}}$, we can compute both $\mu_{\gamma\gamma}$ and the coupling product $(yy^c)$. Using $y,y^c$ as initial conditions, we run the theory up in scale. The renormalization group equations (RGEs) are given in App.~\ref{app:vs}. 
In the left panel of Fig.~\ref{fig:stab}, we plot bands of constant $\mu_{\gamma\gamma}$ (pink) in the $(m_{L_1},m_{L_2})$ plane. The width of the bands is obtained by varying $\Delta_m$ (see Eq.~(\ref{eq:mspl})) from zero to one. We also plot bands of $\Lambda_{UV}$ in gray. In dark we tune $y=y^c$ and in pale we set $y=2y^c$ (the same result is obtained for $y^c=2y$).  

Only a very small window of masses, 100~GeV$<m_{L_1}<$115~GeV and $m_{L_2}\gsim430$~GeV, is compatible with $\mu_{\gamma\gamma}>1.5$ and $\Lambda_{UV}>10$~TeV. Even allowing for $\Lambda_{UV}=1$~TeV, the maximal lighter state mass compatible with $\mu_{\gamma\gamma}>1.5$ is bounded by $m_{L_1}\lesssim140$~GeV. The maximum possible value of $\mu_{\gamma\gamma}$ compatible with $\Lambda_{UV}>1$~TeV is $\approx1.8$.

One can repeat the same exercise for larger $\mathcal{N}$. For instance, for $\mathcal{N} = 4$, we find that allowing for $\Lambda_{UV}=10$~TeV, the maximal lighter state mass compatible with $\mu_{\gamma\gamma}>1.5$ is bounded by $m_{L_1}\lesssim200$~GeV.

\paragraph{Vector doublets + triplet (``wino-higgsino"):  $\psi,\psi^c\sim(1,2)_{\pm\frac{1}{2}},\;\;\chi\sim(1,3)_{0}.$}
We identify $\chi$ and $\chi^c$; the Lagrangian leading to (\ref{eq:MM}) is
\ba-\mathcal{L}=m_\psi\psi\psi^c+\frac{1}{2}m_\chi\chi\chi+\sqrt{2}yH\psi\chi+\sqrt{2}y^cH^\dag\psi^c\chi+cc.\ea
As in the ``vector-like lepton" model, there are two charged Dirac fermions with masses $m_l, m_h$ ($m_l < m_h$). Again, the relevant RGEs are given in App.~\ref{app:vs}. The results are depicted in the right panel of Fig.~\ref{fig:stab}. The bounds on $\mu_{\gamma\gamma}$ are somewhat more severe than for the previous example, with $\Lambda_{UV}\geq10$~TeV and $\mu_{\gamma\gamma}\geq1.5$ only possible for $m_l < 105$~GeV. Allowing for $\Lambda_{UV}=1$~TeV, the maximal lighter state mass compatible with $\mu_{\gamma\gamma}>1.5$ is bounded by $m_{l}\lesssim130$~GeV. The maximum possible value of $\mu_{\gamma\gamma}$ compatible with $\Lambda_{UV}>1$~TeV is $\approx1.75$.

The ``wino-higgsino" example also coincides with SUSY, where $\chi$ and $\psi,\psi^c$ play the role of the wino and higgsinos. We show the SUSY result by green dashed lines (achieved by varying $\mu, M_2$) in the right panel of Fig.~\ref{fig:stab}. In this case $y,y^c$ are limited by the gauge couplings $g\sin\beta,g\cos\beta\lsim0.5$, so the diphoton effect is modest, $\mu_{\gamma\gamma}\lsim1.2$.\\

The choice $y=y^c$ maximizes the value of $\Lambda_{UV}$ for a fixed $\mu_{\gamma\gamma}$. This amounts to some fine-tuning of parameters: given $\mu_{\gamma\gamma}$, the product $(yy^c)$ is essentially fixed and so the cut-off scale is very sensitive to mismatch $y\neq y^c$, as the Higgs quartic runs with $(d\lambda/dt)\propto y^4+y^{c4}$. This result is clear in Fig.~\ref{fig:stab}, where, already for mild splitting $y=2y^c$, the pale gray band of $\Lambda_{UV}=10$~TeV excludes $\mu_{\gamma\gamma}\gsim1.4$. 

Admitting large charge $Q^2>1$ or considering multiple copies of fermions, $\mathcal{N}>1$, would increase the cut-off $\Lambda_{UV}$ for a fixed $\mu_{\gamma\gamma}$ and fermion mass. 
The fact that $\Lambda_{UV}$ rises with $\mathcal{N}$ can be understood as follows. If we rescale $\mathcal{N}$ at fixed $\mu_{\gamma\gamma}$ and mass $m_{L_1}$, the weak-scale initial condition for the Yukawa RGE changes roughly as $y_0^2\to(y_0^2/\mathcal{N})$. Keeping only the $y$ terms in the Yukawa RGEs we have $d(\mathcal{N}y^2)/dt\propto(\mathcal{N}y^2)^2$; hence the running coupling $(\mathcal{N}y^2)$ is approximately invariant under $\mathcal{N}$ rescaling. Including only the $y^4$ contribution in the running of the Higgs quartic $\lambda$, we now have $d\lambda/dt\propto(\mathcal{N}y^2)^2/\mathcal{N}$. As a result, the cut-off scale shifts roughly as $\Lambda_{UV}\to\Lambda_{UV}^\mathcal{N}$. 
In Fig.~\ref{fig:stab2} we repeat our exercise of Fig.~\ref{fig:stab} with $\mathcal{N}=2$ identical copies of vector like fermions. The maximal lightest fermion mass compatible with $\mu_{\gamma\gamma}=1.5$ is somewhat larger than for $\mathcal{N}=1$, but still not larger than $\sim150$~GeV for $\Lambda_{UV} \gsim 10$ TeV. 

Instead of doubling our basic ``vector-like lepton" model, it is arguably more economical to add only vector-like $SU(2)$ singlets, or doublets, but not both. It is straightforward to show, however, that the vacuum stability constraint in this case is more severe than for $\mathcal{N}=2$ copies of the full set-up. The reason is that the Yukawa and Higgs quartic RGEs in the two possibilities are the same, up to an un-important difference in the SM gauge beta functions, while the diphoton enhancement from the three resulting charged Dirac eigenstates cannot exceed that from the four eigenstates of $\mathcal{N}=2$. A similar conclusion applies if we extend the ``wino-higgsino" model by adding either extra triplets or extra doublets but not both. 

Finally we return briefly to the possibility, explored in Fig.~\ref{fig:colornocolor} and the corresponding discussion, that the diphoton enhancement is produced by colored particles. There, we argued in general that the colored solution inevitably involves fine tuning, regardless of the spin of the particle. For colored fermions, this possibility is also strongly constrained by vacuum stability, as it requires very large Yukawa couplings. Indeed, calculating the RGEs for a generation of vector-like up-type quarks\footnote{The field content we consider is $\psi\sim(3,2)_{\frac{1}{6}},\;\psi^c\sim(\bar 3,2)_{-\frac{1}{6}},\;\chi\sim(\bar 3,1)_{-\frac{2}{3}},\;\chi^c\sim(3, 1)_{\frac{2}{3}}$. See~\cite{Choudhury:2001hs, Morrissey:2003sc, Dawson:2012di} for electroweak precision constraints on this field content, in the context of modified Higgs couplings.}, we find that imposing $\Lambda_{UV}>1$~TeV implies $\mu_{\gamma\gamma}<1$.

To conclude, a diphoton enhancement $\mu_{\gamma\gamma}=1.5$ through a minimal vector-like set of fermions requires a light charged state with mass below $115$~GeV, even when we allow a very low cut-off scale $\Lambda_{UV}=10$~TeV for the theory and judiciously tune the parameters to maximize the effect by setting\footnote{See Eqs.~(\ref{eq:MM}-\ref{eq:mspl}) and the discussion between them for the definition of $y,\,y^c,\,\Delta_m$ and $\phi$.} $\phi=0,\,\Delta_m=0$, and $y=y^c$. Relaxing the parameter tuning slightly brings us down to the LEPII bound, excluding the model or, at best, implying that the numerical value of the mass is tuned. Extending the set-up to $\mathcal{N}=2$ identical copies of vector-like fermions allows for slightly less precise parameter tuning (though the number of tuned parameters is doubled), but the lightest fermions must still lie below $\sim150$~GeV for $\Lambda_{UV}\gsim10$~TeV. Even allowing $\mathcal{N}=4$ identical copies of vector-like fermions, the upper bound of the lightest fermions' masses only slightly increases to $\sim200$~GeV for $\Lambda_{UV}\gsim10$~TeV. 

\begin{figure}[!h]\begin{center}
\includegraphics[width=0.45\textwidth]{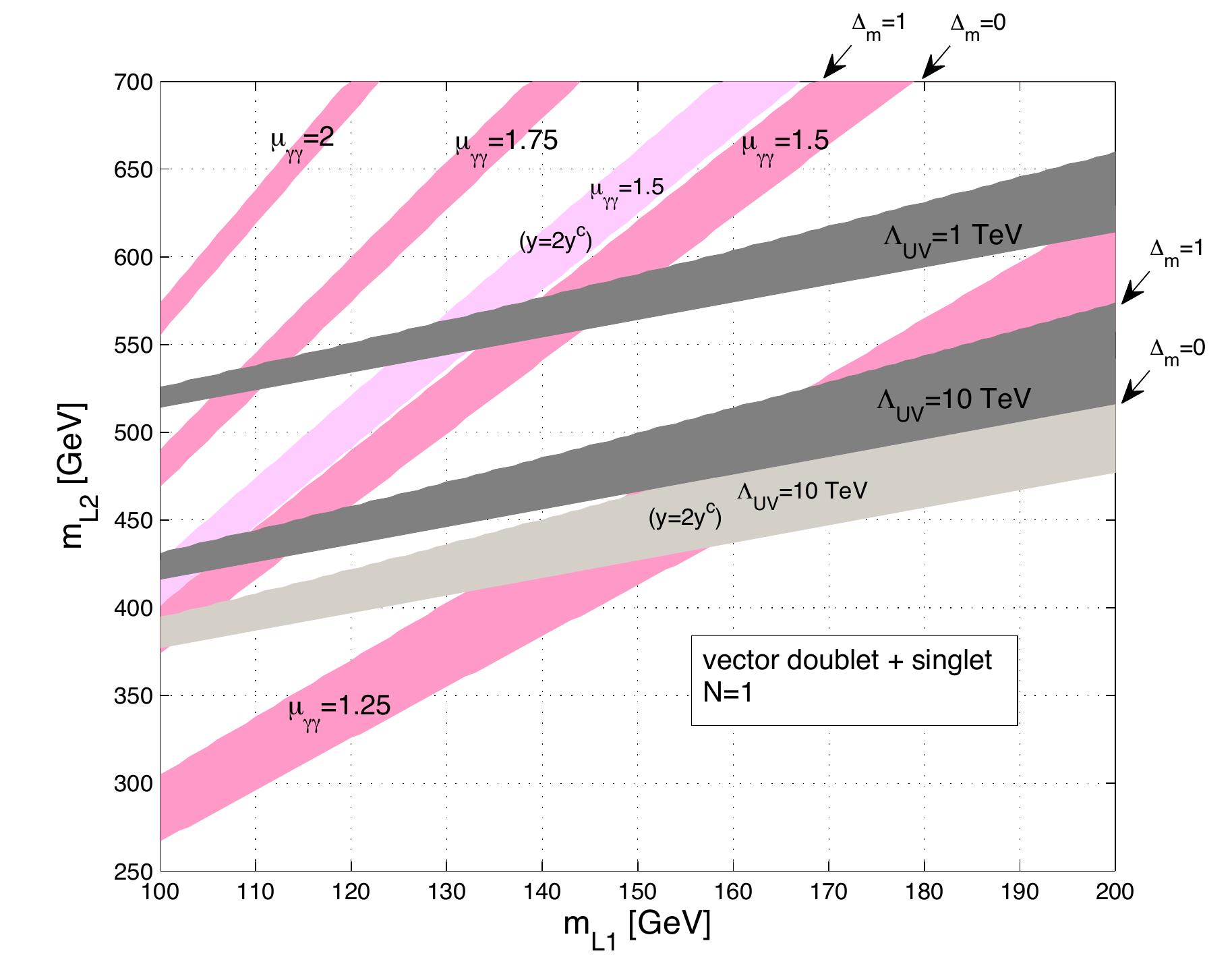}\quad
\includegraphics[width=0.45\textwidth]{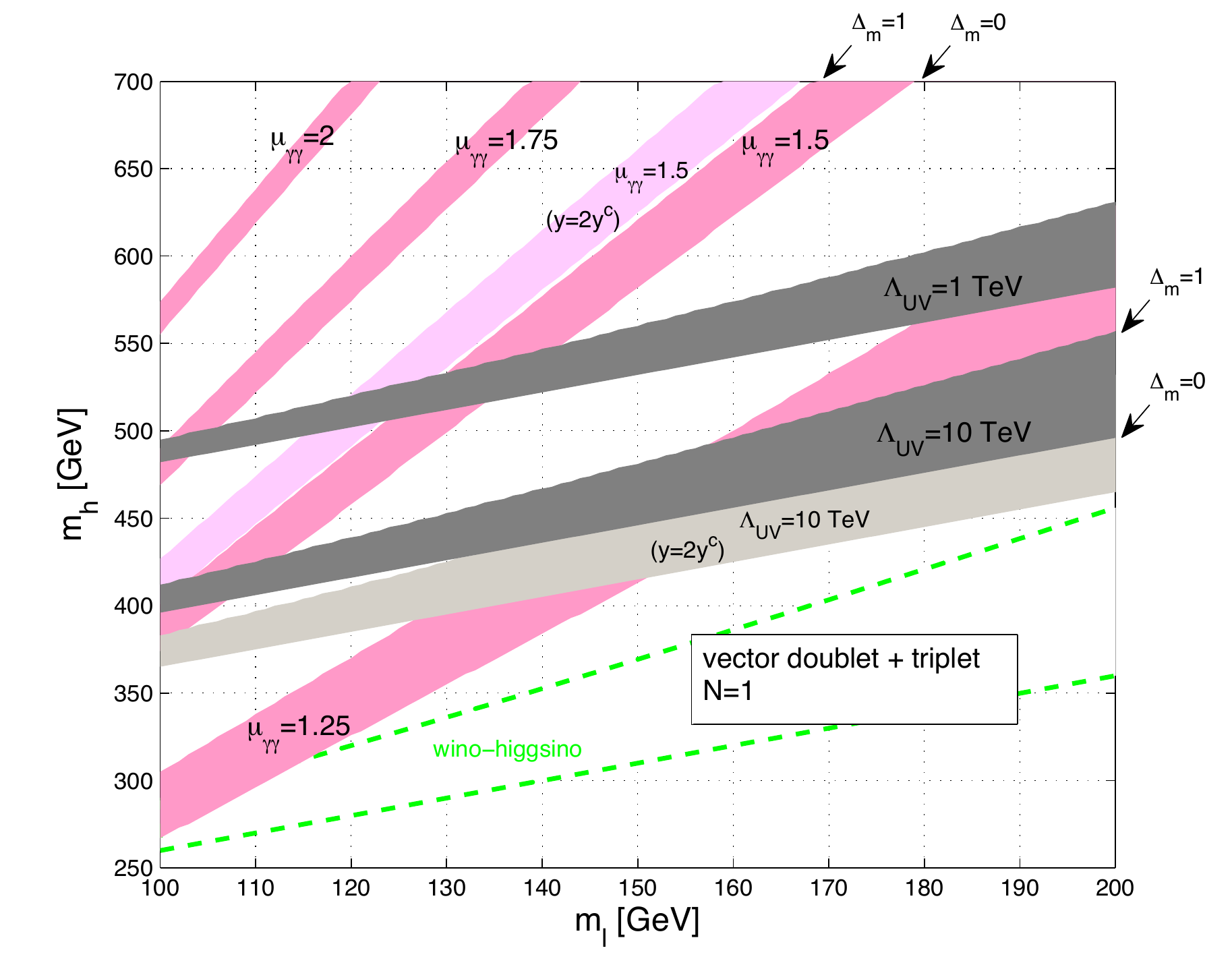}  
\end{center}
\caption{Left: ``vector-like lepton" model. Right: ``wino-higgsino" model. The horizontal and vertical axes correspond to the light and heavy mass eigenvalues, respectively. Pink bands denote the diphoton enhancement $\mu_{\gamma\gamma}$. Gray bands denote the vacuum instability cut-off $\Lambda_{UV}$. Dark  is for $y=y^c$; pale is for $y=2y^c$. The width of the bands (for both $\mu_{\gamma\gamma}$ and $\Lambda_{UV}$) correspond to varying the electroweak-conserving mass splitting term $\Delta_m$ (see Eq.~(\ref{eq:mspl})) from zero to one. Green dashed band, on the right, denotes the SUSY wino-higgsino scenario.}
\label{fig:stab}
\end{figure}%

\begin{figure}[!h]\begin{center}
\includegraphics[width=0.45\textwidth]{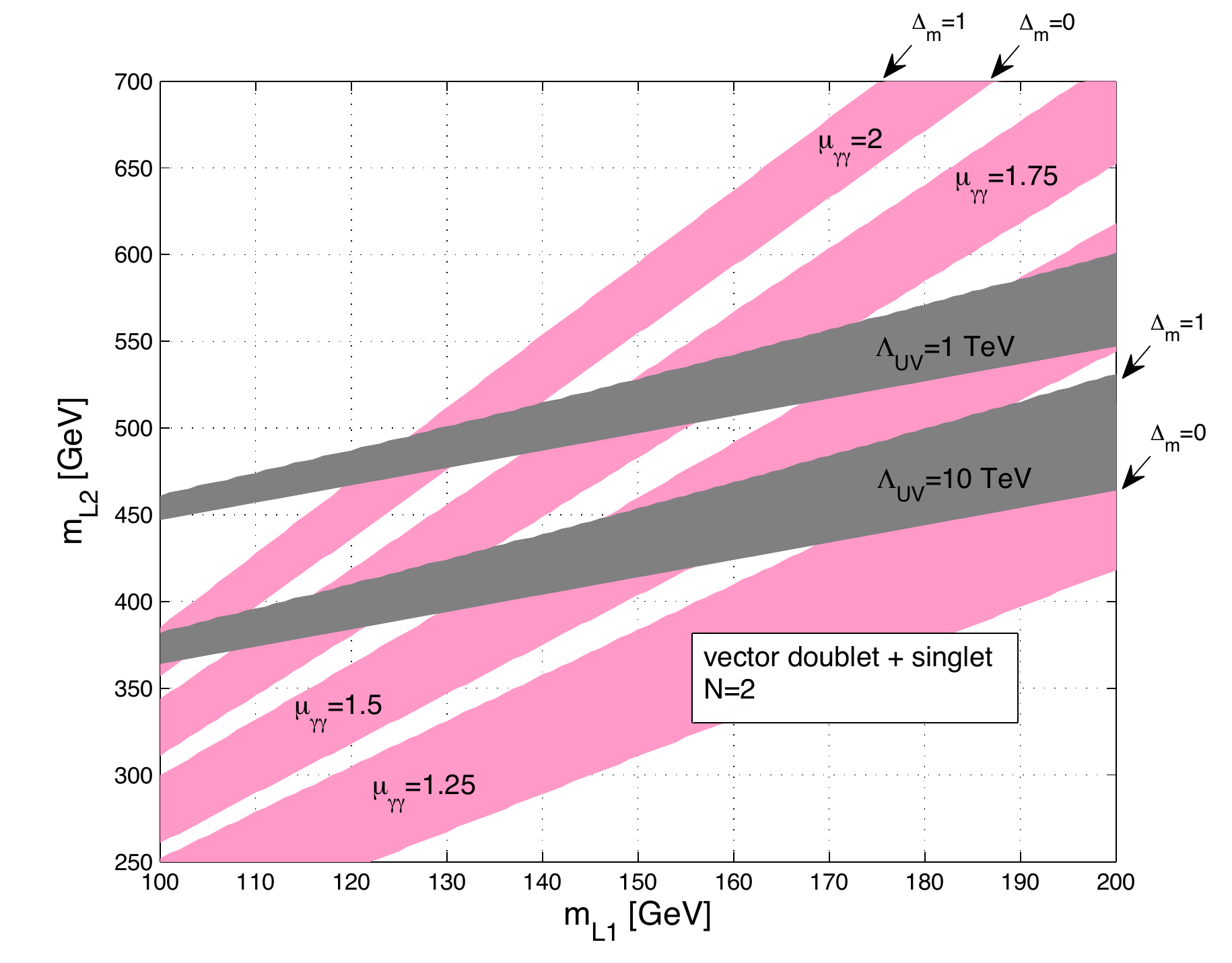}\quad
\includegraphics[width=0.45\textwidth]{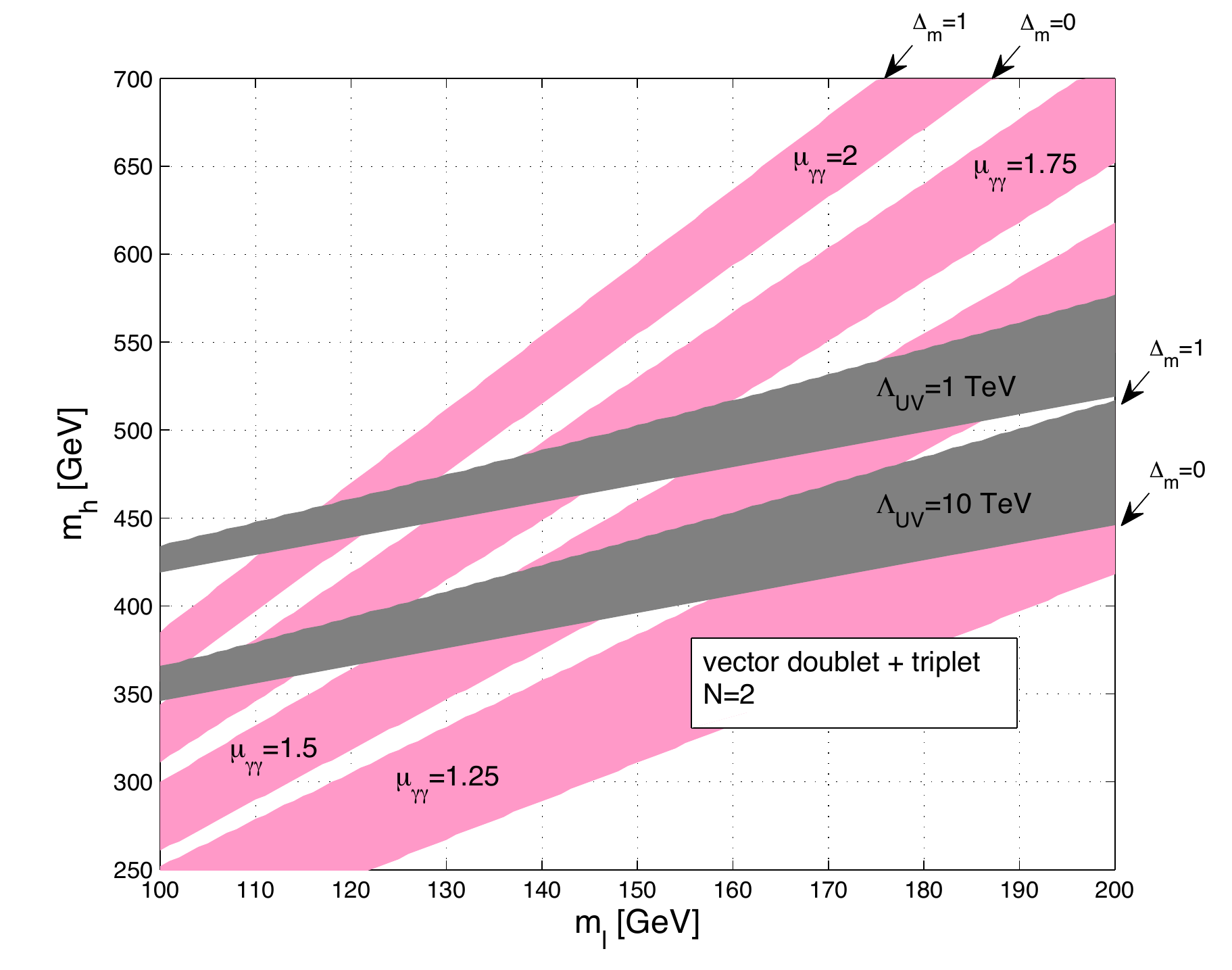}  
\end{center}
\caption{Same as Fig.~\ref{fig:stab}, but for $\mathcal{N}=2$ copies of vector like fermions.}
\label{fig:stab2}
\end{figure}%

\section{Collider signals and electroweak constraints}
\label{sec:col}
The light charged fermions discussed in the previous section are produced through electroweak processes with appreciable rates at hadron colliders. In this section we consider constraints and detection prospects from current and upcoming searches, assessing charchteristic detection channels and providing rough estimates of the experimental sensitivity. We stress that our analysis is simplistic, and can by no means replace a full-fledged collider study. Nevertheless, our estimates provide solid motivation and concrete guidelines for a more dedicated study in the future, should the diphoton enhancment be confirmed by upcoming data.
We limit the discussion to a single set of vector-like fermions. We use the notation of our ``vector-like lepton" example, for simplicity, but most of the discussion also applies to the ``wino-higgisino" model. 

The most important production mode for $\mathcal{N}=1$ is $pp \to L_1^+ L_1^-$. 
To calculate the production cross sections we use the FeynRules package~\cite{Christensen:2008py} interfaced with MadGraph 5~\cite{Alwall:2011uj}. In the left panel of Fig.~\ref{fig:xsec} we plot $\sigma(pp\to L_1^+L_1^-)$, in the ``vector-like lepton"  model, as a function of the lightest charged state mass, setting $y=y^c$ and $\Delta_m=0$. Other cross sections are generically smaller, because of the mass gap that is required to enhance the Higgs diphoton coupling. For example, for $m_{L_1} = 100$~GeV, obtaining $\mu_{\gamma\gamma} = 1.5$ requires $m_{L_2} \geq 368$~GeV and $m_{N} \geq 234$~GeV, with $\sigma (pp \to L_1^\pm N) \approx 70$ fb, $\sigma (pp \to N N) \approx 29$ fb and $\sigma (pp \to L_1 L_2) \approx 5$ fb at the LHC with $\sqrt{s}=$7 TeV. 

In the right panel of Fig.~\ref{fig:xsec} we plot the cross section of the lightest charged state pair production in the ``wino-higgisino"  model. The cross section is much larger  compared to the ``vector-like lepton" case. The reason is that in the $y=y^c,\,\Delta_m=0$ limit, where the singlet and doublet components are maximally mixed, the lightest charged state coupling to the $Z$ boson is suppressed by a small factor $(4 \sin \theta_W^2 -1) \approx 0.08$. Thus $pp(\bar{p}) \to L_1^+ L_1^-$ mainly goes through a photon. The cross section grows away from the $y=y^c,\, \Delta_m=0$ limit, where the doublet component of the lightest state can be increased (at the cost of reducing the Higgs diphoton coupling), and the suppression is absent in the ``wino-higgsino" model. Thus the left panel of Fig.~\ref{fig:xsec} gives the rock bottom lower limit for the cross section expected in our scenario, while the right panel gives a rough upper limit.

\begin{figure}[!h]\begin{center}
\includegraphics[width=0.45\textwidth]{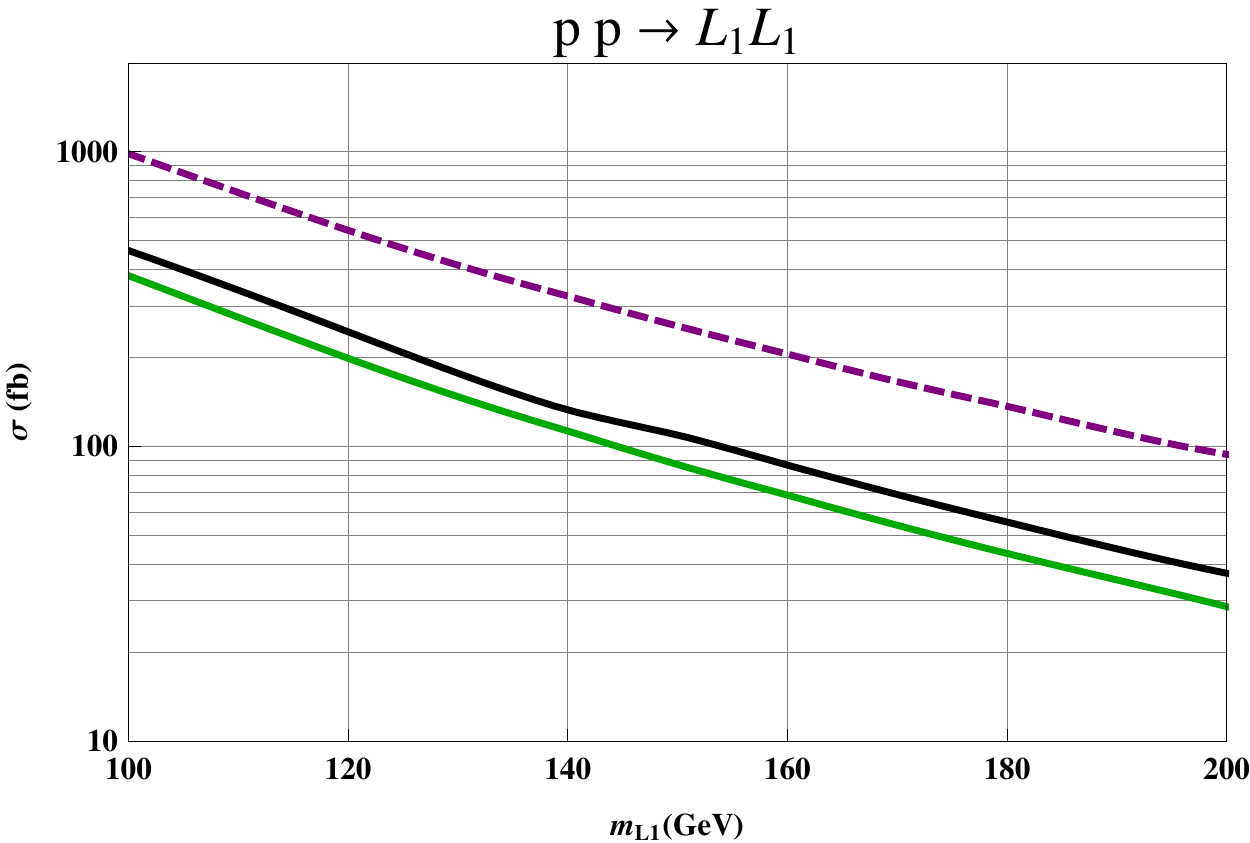} \quad
\includegraphics[width=0.45\textwidth]{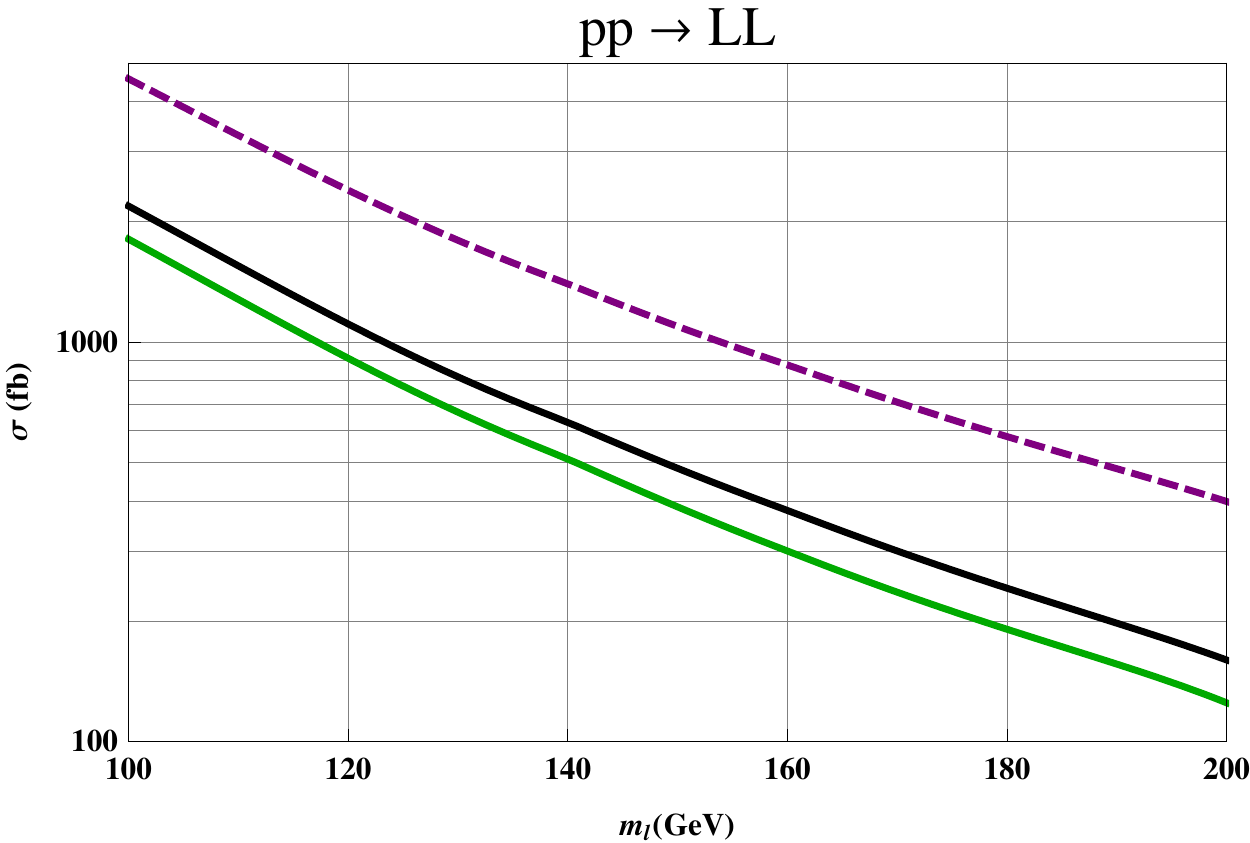}
\end{center}
\caption{Left: $\sigma(pp\to L_1^+L_1^-)$ as a function of the lightest charged state mass in the ``vector-like lepton" model at the LHC7 (green, bottom), LHC8 (black, middle) and LHC14 (purple, top). Right: same, for the ``wino-higgsino" model.}
\label{fig:xsec}
\end{figure}%

In our models, taken as they are, the lightest charged fermion is stable. For the masses of interest, however, this possibility is already excluded by searches for heavy stable charged particles (HSCPs)~\cite{Chatrchyan:2012sp}. It is easy to reach this conclusion by looking at the excluded cross section in the direct stau production case. The rather model independent cuts on the heavy particle $p_T$, time of flight and energy loss by ionization should retain a comparable efficiency on  our signal. We therefore consider two modifications of the minimal set-up:
\paragraph{{\bf(A)}} It is always possible to add one or more extra SM singlets $n$ (``sterile neutrino" or "bino" in the ``vector-like lepton" and ``wino-higgsino" models, respectively), with Yukawa couplings $\mathcal{L}=-y_nH^\dag \psi n-y_n^cH\psi^cn+cc$ and mass term $\frac{1}{2}m_n n n$. 
Mixing with the $SU(2)$ component, $N$, makes room for a neutral state, $n_1$, with mass below that of the charged $L_1$. This opens up the decay mode $L_1 \to W^{(*)} n_1$, where the $W$ boson can be on- or off-shell depending on the mass splitting between $L_1$ and $n_1$. 
\paragraph{{\bf (B)}} Alternatively, a small mass mixing with the SM leptons would induce decays such as $L_1 \to Z l(\tau)$ and $L_1 \to W \nu$, where $l \equiv e, \mu$ and $\nu \equiv \nu_e, \nu_\mu, \nu_\tau$. Constraints on the flavor changing processes $\mu \to e \gamma$ and $\tau \to e \gamma$ limit the mixing angles to $|U_{eL}U_{\mu L}| < 10^{-4}$ and $|U_{eL}U_{\tau L}|<10^{-2}$~\cite{Nakamura:2010zzi}. 
Additional constraints arise from LEP measurements of the $Z$ widths to leptons~\cite{Nakamura:2010zzi}, that are roughly known to $\sim1$ part in $10^4$ for each of the three generations. We  thus require conservatively $|U_{iL}| \lesssim 10^{-2}$ for $i=e, \mu, \tau$. Note that as long as a mixing angle is bigger than $\sim10^{-4}$, the decay is prompt\footnote{Searches for displaced vertices and long lived particles decaying inside the detector are currently ongoing at the LHC (see for example \cite{cmsvertex}) and were performed at the Tevatron \cite{Acosta:2005np, Scott:2004wz, Abazov:2009ik, Abazov:2008zm, Abazov:2006as}, but there is still no systematic coverage of all the possible lifetimes and final states. We will ignore this possibility in what follows, even though experimentally it is intriguing and could be the subject of a dedicated study.}.

In case {\bf (A)}, the main signature is the pair production of  two charged particles decaying to $W^{(*)}W^{(*)}$+MET, depicted in the left panel of Fig.~\ref{fig:processes}. Recently an ATLAS analysis targeting final states with two leptons and MET was released~\cite{ATLASos}. After the full selection in~\cite{ATLASos}, in the $m_{T2}$ signal region an efficiency ranging between $1\%$ and $7\%$ is observed for a signal consisting of chargino pair production, while the measured range for slepton pair production is lower. 
We take the same efficiency on our signal as that measured in the chargino case. Note that this is only an order of magnitude estimate, as the decay chains are not identical.

The limit on the cross section (that takes into account the $WW$ leptonic branching ratio) from the ATLAS $m_{T2}$ signal region (all flavor combined channel) is $42$ fb. This should be compared with the cross sections in Fig. \ref{fig:xsec} multiplied by the efficiency assumed above. We find that in the ``vector-like lepton" case there is not enough sensitivity to probe cross sections comparable to ours, while for large enough mass splittings between $L_1$ and $n_1$ (efficiency $\sim 7\%$) we can already exclude the interesting mass range in the ``wino-higgsino" model ($m_{l}\lesssim 140$~GeV) and it is likely that the LHC will be sensitive to the ``vector-like lepton" model by the end of the year ($\sigma \times \epsilon\approx 31(7)$fb for $m_{L_1}=100(140)$~GeV).

Note that the minimum mass splitting between chargino and LSP considered in the analysis above is always greater than 75 GeV. In our
case, decreasing the splitting between $m_{L_1}$ and $m_{n_1}$, MET and $m_{T2}$ cuts quickly loose efficiency and eventually even final state leptons become too soft to be triggered. The only experimental handles in this case are monojet and monophoton + MET searches~\cite{CDF, cmsmonojet, cmsmonophoton, atlasmonojet, atlasmonophoton}, that can also detect the pair production of the lightest neutral state. Current searches are beginning to probe colored particle production cross sections for masses in the few hundred GeV range~\cite{Belanger:2012mk, Dreiner:2012gx}, not having yet sensitivity to electroweak production. Translating current limits on new signals is a non-trivial task in the monojet case, due to the large uncertainties on the simulation of ISR \cite{Dreiner:2012gx} and it is even harder to make predictions for the near future. However it was estimated that the discovery reach of the 14 TeV LHC is about 200 GeV for a gaugino LSP and that masses around 120 GeV can be probed already with 10 fb$^{-1}$ if systematic uncertainties are kept under control~\cite{Belanger:2012mk, Ibe:2006de, Giudice:2010wb}.
Monojet estimates must be taken with caution, for the reasons mentioned above, but it was also shown that at the LHC an ISR jet has often a companion in the event. Therefore the results from a second channel with two jets and MET can be combined with the monojet searches to further increase the sensitivity. Attempts in this direction have already been made and the CMS ``razor" analysis \cite{cmsrazor} was shown to have a comparable sensitivity to dark matter production to that of monojet searches~\cite{Dreiner:2012gx, Fox:2012ee}.

In case {\bf (B)}, several different processes lead to multi-lepton final states with little hadronic activity. This scenario is depicted in the middle panel of Fig.~\ref{fig:processes}. The CMS multilepton search~\cite{Chatrchyan:2012ye} is currently the most sensitive to final states with low MET, and can already exclude a large fraction of the relevant parameter space. Here, for simplicity, we consider a few decay modes in single exclusive channels. If we take, for example, $\mathrm{BR}(L_1 \to Z +l)\approx100 \%$, and assume a flat $70\%$ efficiency times acceptance for each of the four leptons\footnote{From \cite{ContrerasCampana:2011aa} we get an efficiency of the kinematical cuts $\sim 0.87$. Taking into account the finite acceptance (somewhat optimistically) we obtain the final 0.7 \cite{thomas}. Notice that this is a huge simplification of the experimental set-up that does not even distinguish between electrons and muons, and is thus only intended to give an order of magnitude estimate.}, then we find that the region $m_{L_1} \subset (100 - 120)$ GeV can be excluded already in the ``vector-like lepton" model, while the limit extends to $\approx 180$~GeV in the ``wino-higgsino" case. This estimate was made using a standard $\mathrm{CL}_s$ technique described in App.~\ref{sec:limits} from a single channel $4l$ with MET $<50$~GeV and $H_T<200$~GeV and a Drell-Yan lepton pair from a $Z$ decay. 
The limits are slightly weaker for $L_1 \to Z + \tau$. In this case we are not yet sensitive to the ``vector-like lepton" model, while we are sensitive to masses up to $\approx 130$~GeV in the ``wino-higgsino" one. Again this estimate was obtained by looking at a single channel: $3l+1\tau_h$\footnote{Assuming an hadronic tau identification efficiency, for the HPS algorithm used in the CMS paper, $\epsilon_{\tau_h}=0.35$ \cite{2012JInst.7.1001C, PFT-10-004, PFT-08-001, thomas} and the same $0.7$ efficiency as before for any extra lepton.} with MET $<50$~GeV and $H_T<200$~GeV and a Drell-Yan lepton pair from a $Z$ decay. It is clear that the rest of the relevant parameter space can easily be covered by the end of the year and that, combining different channels and possibly results from the two experiments, the sensitivity would be increased, covering also more generic scenarios in which the branching ratio to these final states is not exactly one. 

In summary, for $\mathcal{N}=1$, an $L_1$ decaying to SM leptons is either already ruled out or within reach of the 8 TeV LHC. If, instead, $L_1 \to W^* n_1(\nu)$ dominates, the relevant final state is $WW +$ MET from $L_1L_1$ production, which is still unconstrained for the ``vector-like lepton" model, but also within reach of the 8 TeV LHC. In the worst case, when $L_1$ and $n_1$ are nearly degenerate in masses, the monojet searches will be able to probe the relevant parameter space at the 14 TeV LHC. In the latter case, other interesting channels, especially for $\mathcal{N}>1$, would be the $WWW$+MET and, to a lesser extent, $WZ$+MET final states arising from the production of $L_1N$ as depicted in the right panel of Fig.~\ref{fig:processes}. 
Dedicated analyses, beyond the scope of this paper, would improve the current sensitivities for some of the channels\footnote{It is sufficient to think about possible three-lepton resonance searches or monojet searches with the additional requirement of soft leptons in the final state \cite{Giudice:2010wb}.}. 
%
For our purpose here it suffices to show that if the enhancement of the $\gamma\gamma$ rate will be confirmed and an un-natural theory is responsible for it, then we expect the new fermions involved to be detected in the next few years, or even months.

\begin{figure}[!h]\begin{center}
\includegraphics[width=0.26\textwidth]{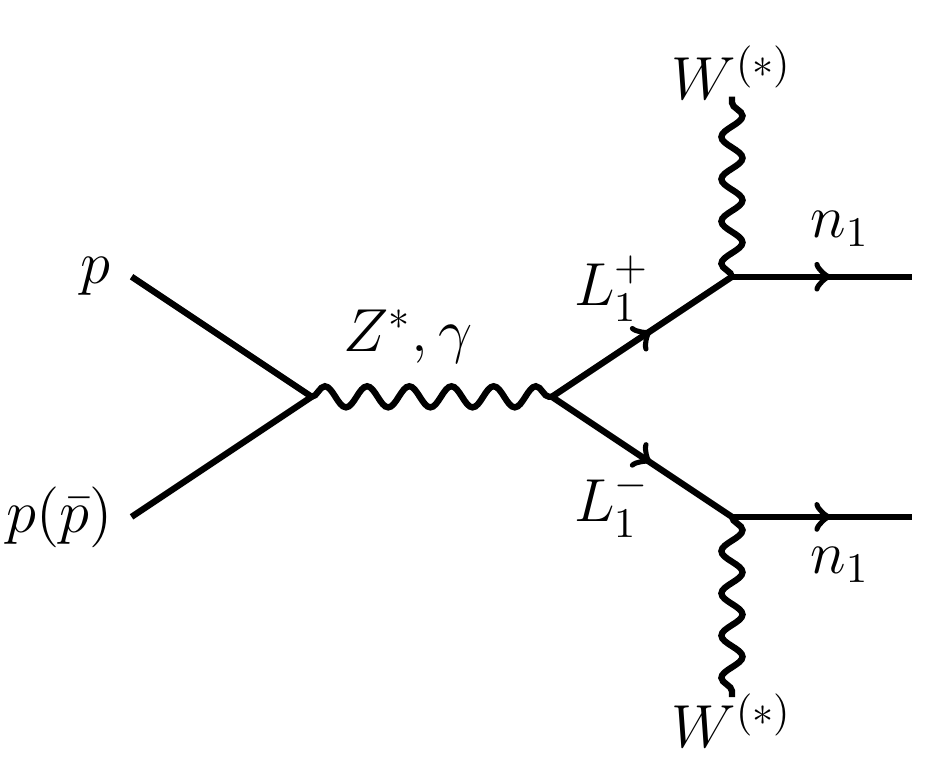} \quad  
\includegraphics[width=0.28\textwidth]{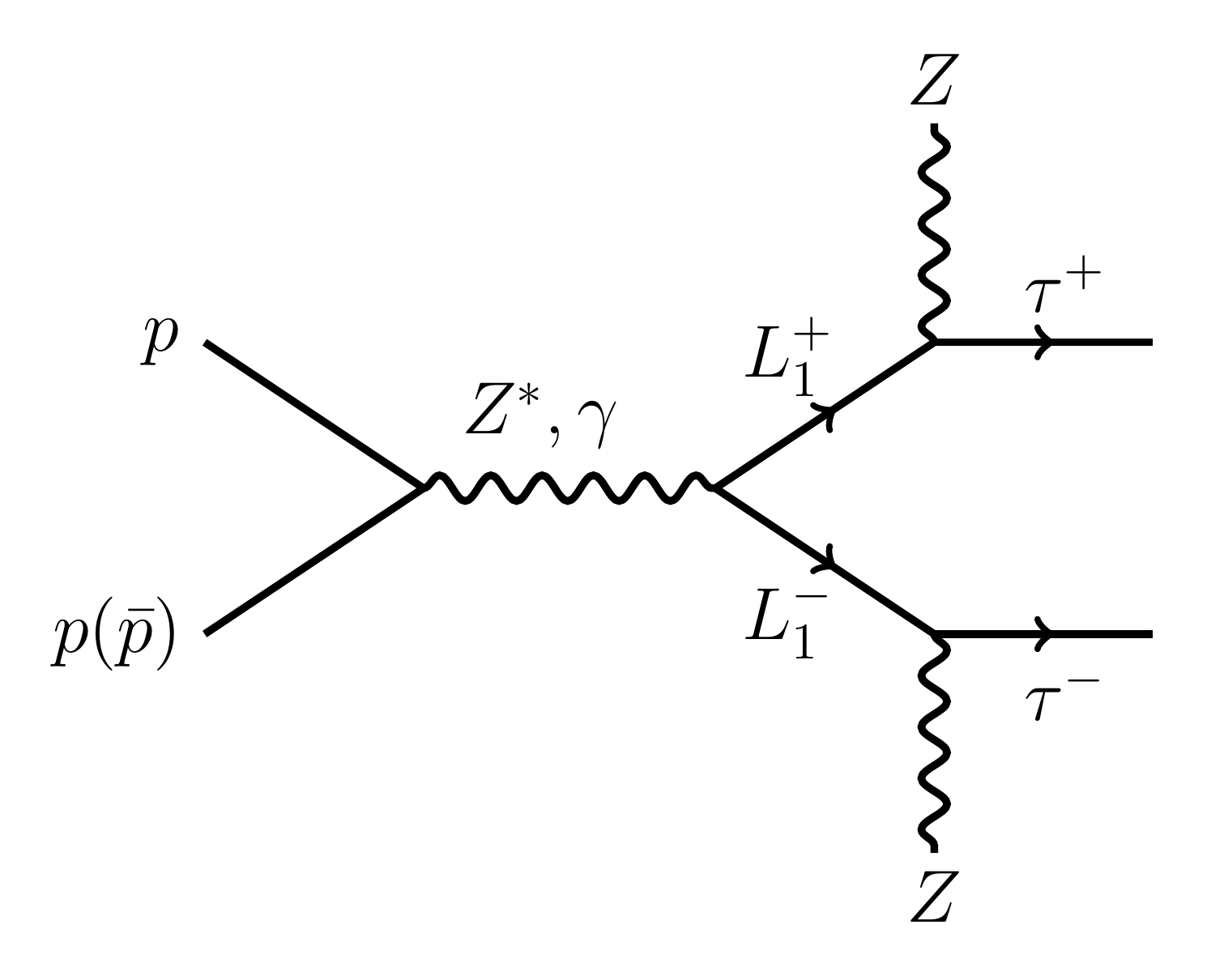} \quad
\includegraphics[width=0.3\textwidth]{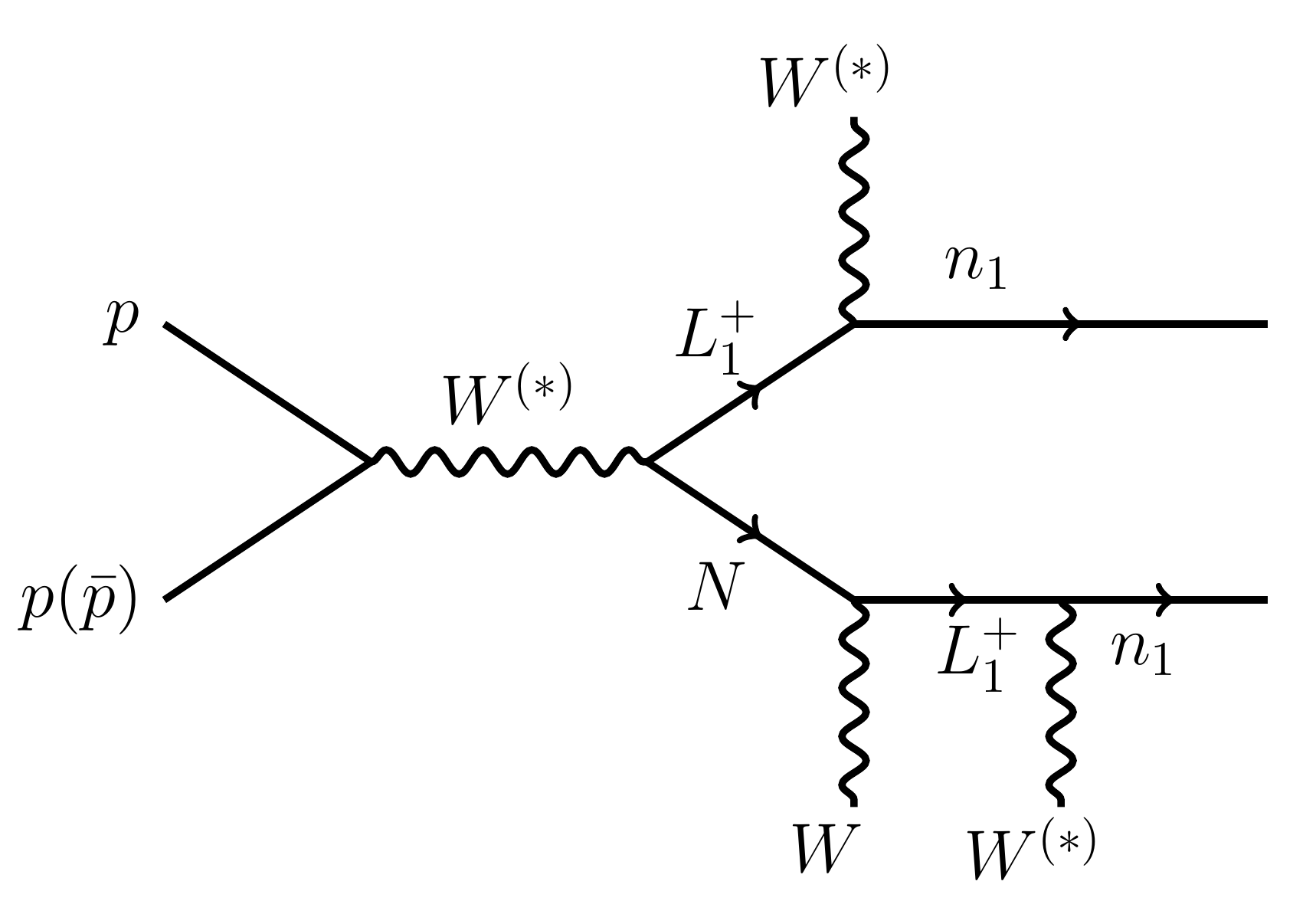} 
\end{center}
\caption{Feynman diagrams for new fermion production and decay.}
\label{fig:processes}
\end{figure}%

Finally, in addition to direct searches, light non-singlet fermions are constrained indirectly by electroweak precision tests (EWPTs), especially so given the need for a large electroweak breaking mass to affect $\mu_{\gamma\gamma}$. Indeed, specializing to the ``vector-like lepton" example\footnote{We expect similar results to hold for the ``wino-higgsino" model, as can be deduced e.g. from~\cite{Martin:2004id}.}, in the minimal field content specified by Eq.~(\ref{eq:doubL}), we find that $\mu_{\gamma\gamma}>1.5$ comes along with a sizable $T$ parameter, whereas $\mu_{\gamma\gamma}>1.75$ would be firmly excluded. Nevertheless, the tension with EWPTs can be tuned away by means of additional free model parameters. For instance, mixing with a neutral singlet $n$, as discussed earlier, produces an opposite contribution to $T$ that can bring the model back to life even for $\mu_{\gamma\gamma}=2$. Since this counter effect relies, again, on sizable Yukawa couplings $y_n,y_n^c$, it comes at the cost of lowering somewhat further the instability cut-off $\Lambda_{UV}$. 

In Fig.~\ref{fig:ST} we illustrate this behavior by computing $S$ and $T$, following~\cite{Maekawa:1994yd} and performing the EWPT fit for $m_h=125$~GeV~\cite{Baak:2011ze}. In the left panel, we indicate with a green shaded area the 95\%CL EWPT exclusion region in the $(m_{L_2},x_n)$ plane. Here, $x_n$ is defined in analogy with Eq.~(\ref{eq:mspl}) as $x_n^2=(2y_ny_n^cv^2/m_{n_1}^2)$, where $m_{n_1}$ is the lighter neutral state mass, and $m_{L_2}$ is the mass of the heavier charged state. We set $m_{L_1}=100$~GeV, $y=y^c,\,y_n=y_n^c$ and $m_\psi=m_\chi=m_n$. Also plotted are the diphoton enhancement (pink) and values of $\Lambda_{UV}$ (gray). In reading the plot, note that walking on the horizontal axis towards larger $m_{L_2}$ is equivalent to walking up on the left edge of the left panel of Fig.~\ref{fig:stab}. We see that with some neutral mixing, it is possible to tune away the tension with EWPTs, even for large $\mu_{\gamma\gamma}$. In the right panel of Fig.~\ref{fig:ST} we show on the $S-T$ ellipse three sample model points, marked correspondingly on the left.
\begin{figure}[!h]\begin{center}
\includegraphics[width=0.45\textwidth]{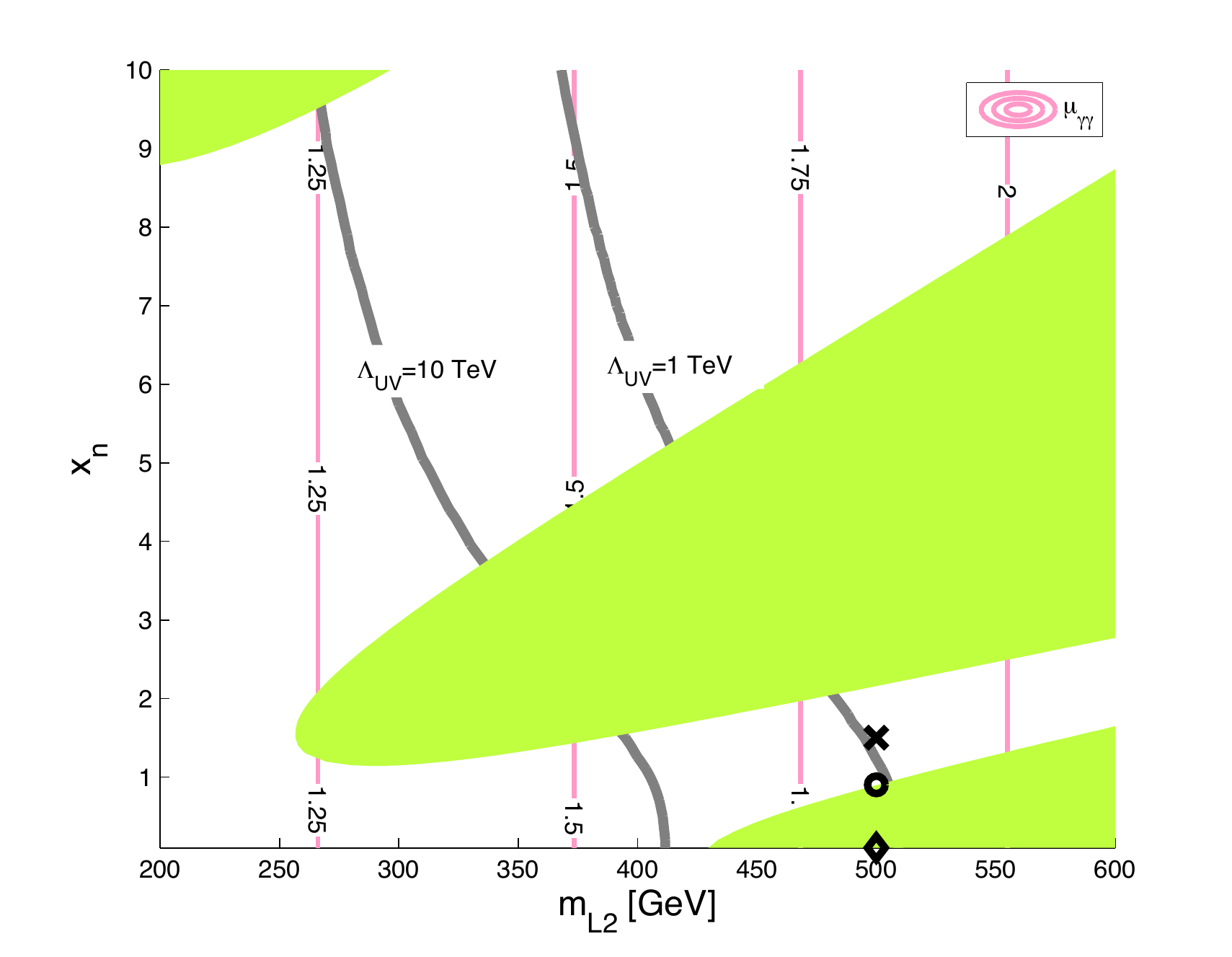}\quad
\includegraphics[width=0.45\textwidth]{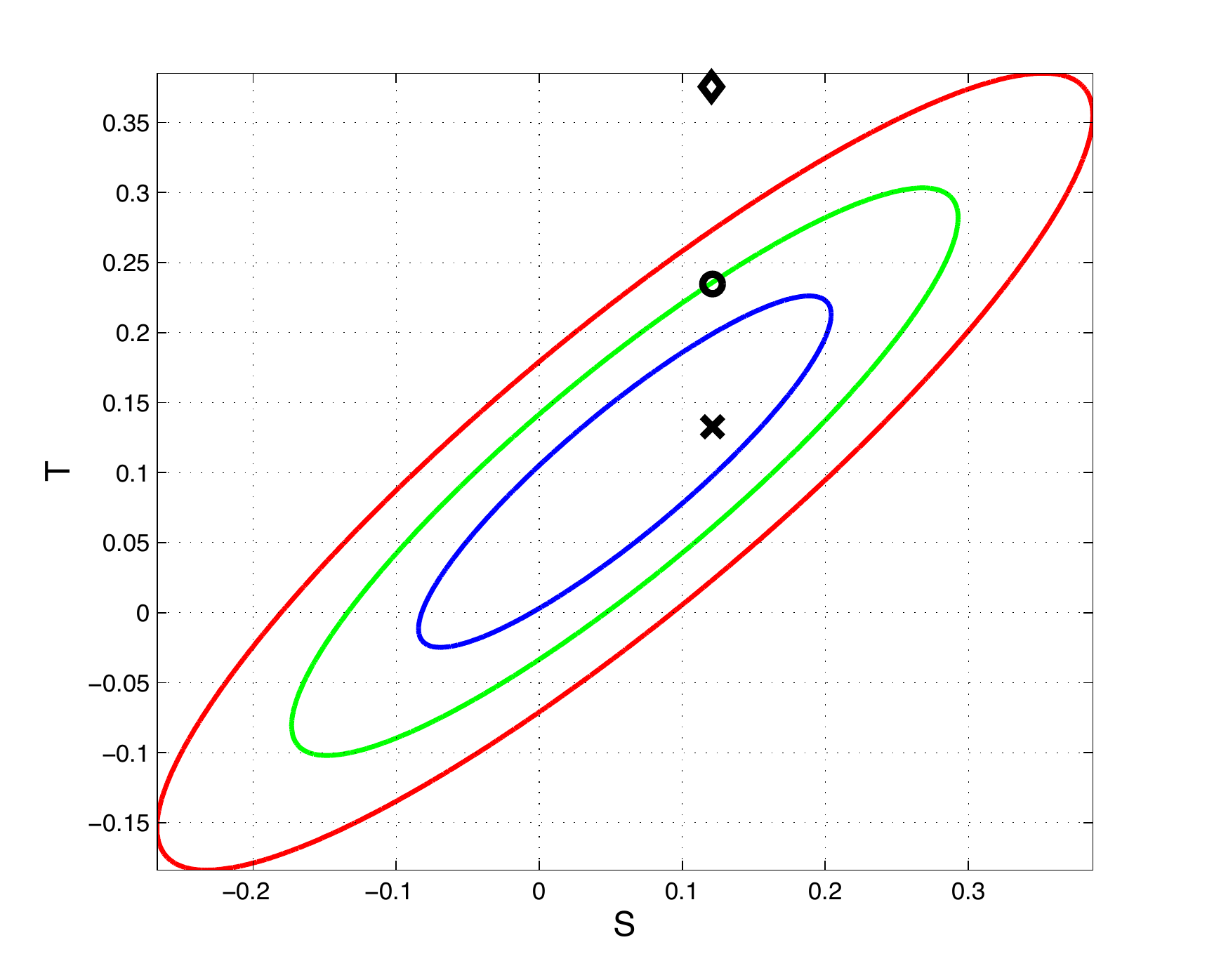}  
\end{center}
\caption{Electroweak constraints for the ``vector-like lepton" model. Left: contours of $\mu_{\gamma\gamma}$ (pink) plotted in the $(m_{L_2},x_n)$ plane, where $m_{L_2}$ is the heavier charged state mass and $x_n$ is defined in analogy with Eq.~(\ref{eq:mspl}) as $x_n^2=(2y_ny_n^cv^2/m_{n_1}^2)$, with $m_{n_1}$ the lighter neutral state mass. Gray lines denote the vacuum instability cut-off $\Lambda_{UV}$. The green filled area is excluded at 95\%CL or more by EWPTs. The lighter charged fermion mass is fixed to $m_{L_1}=100$~GeV. Right: Markers show the model position w.r.t. the $S-T$ error ellipse, for three sample points in the left panel. Blue, green and red lines denote the 68.27\%CL, 95\%CL and 99.73\%CL ranges.}
\label{fig:ST}
\end{figure}%

\section{Discussion and conclusions}
For a single set of new vector-like fermions, with large enough Yukawa
couplings to give an enhancement of $\mu_{\gamma \gamma} = 1.5$, demanding
that the tunneling rate through false vacuum bubbles of size
$\Lambda_{UV}^{-1} \sim(10 ~\rm TeV)^{-1}$ is less than the age of the universe
requires the existence of a new, un-colored, charged fermion lighter than about $115$~GeV. Even with a very low cut-off scale, $\Lambda_{UV}=1$~TeV, an
enhancement of $\mu_{\gamma \gamma} =  2$ is impossible. 

A larger number
$\mathcal{N}$ of fermions allows us parametrically to keep a large enhancement for
$\mu_{\gamma \gamma}$ while ameliorating vacuum stability. It is in
principle possible, though contrived, to get $\mu_{\gamma \gamma} = 1.5$
while deferring the instability scale to $\Lambda_{UV} \sim$ 10 TeV,
though even for $\mathcal{N}=2 (4)$ this requires the fermions to be lighter than 150 (200) GeV. 

Furthermore the cases with large ${\cal N}$ are in great tension with any picture of gauge coupling unification in the UV. Let us look at theories which add vector-like matter to split SUSY. One of the main motivations 
for split SUSY is maintaining supersymmetric gauge coupling unification, but it is easy to see that this feature is lost with a large number of multiplets.
Consider the case where the new vector-like matter is in complete multiplets of $SU(5)$. The ``vector-like lepton" fit inside a full generation + antigeneration; ${\cal N} =1$ of these multiplets are consistent with perturbative gauge coupling unification, but ${\cal N} > 1$ are not. Similarly, the ``wino-higgsino" multiplet can fit in a ${\bf 24}$ of $SU(5)$, and again only ${\cal N} =1$ is (marginally) consistent with perturbative unifcation.
We can even go as far as to consider complete multiplets of $SU(3)^3/Z_3$. 
The multiplet $(3,1,1) + (1,3,1) + (1,1,3) +$ conjugate contains exactly stable uncolored fractionally charged particles; these could give a diphoton enhancement but are forced to be so heavy by HSCP searches that the required Yukawa couplings are too large to be consistent with vacuum stability even for ${\cal N} = 1$ multiplet. The usual matter multiplet $(3, \bar{3}, 1) + (1,3, \bar{3}) + (\bar{3},1,3) +$ conjugate is too large even for ${\cal N} = 1$. Finally we can consider $(8,1,1) + (1,8,1) + (1,1,8)$; this contains both the ``vector-like lepton" and ``wino-higgsinos". But again gauge coupling unification restricts us to having at most one of these multiplets. We conclude that in any reasonable picture preserving perturbative gauge coupling unification, we can have either ${\cal N} = 1$ ``vector-like lepton" or ``wino-higgsino", or at most one of each. 

We thus conclude that even non-minimal un-natural theories at the weak
scale can not explain a large $\mu_{\gamma \gamma}$, unless they have new
charged fermions lighter than about $115 - 150$ GeV. These charged fermions are
so light that in most cases they should be possible to exclude or discover
with the 2012 LHC data. If such light states are not discovered, and at the
same time the large enhancement $\mu_{\gamma \gamma} \sim 1.5 - 2$
persists, then there must be new scalars or gauge bosons far beneath the 10 TeV
scale. The enhanced diphoton rate reported by ATLAS and CMS could be the harbinger of natural electroweak symmetry breaking within reach of the LHC.  Alternately, fine-tuned theories such as split SUSY or any of its variants unambiguously predict that the hint for an enhanced diphoton rate and unaffected $ZZ$ rate in the current data must disappear. 
 
\acknowledgments{We thank Clifford Cheung, Nathaniel Craig, Rouven Essig, Mariangela Lisanti, Josh Ruderman, Sunil Somalwar, Scott Thomas and the CMS multilepton analysis team for useful discussions and Maurizio Pierini for reading and commenting on the draft. N.A.-H. is supported by the DOE under grant DEFG0291ER40654. K.B. is supported by the DOE under grant DEFG0290ER40542. J.F. is supported by the DOE under grant DEFG0291ER40671.}

\begin{appendix}

\section{RGE and determination of the vacuum instability cut-off scale}\label{app:vs}

\paragraph{Vector doublets + singlets (``vector-like lepton").} For our  ``vector-like lepton" scenario, allowing for an additional neutral singlet $n\sim(1,1)_0$ with Yukawa couplings $\mathcal{L}=-y_nH^\dag\psi n-y_nH\psi^cn+cc$, and including $\mathcal{N}$ copies with identical couplings, the relevant RGEs read~\cite{Kribs:2007nz,Ishiwata:2011hr}
\beq
16\pi^2\frac{dy}{dt}&=&y\left(\frac{3}{2}\left(y^2-y_n^2\right)+\mathcal{N}\left(y^2+y^{c2}+y_n^2+y_n^{c2}\right)+3y_t^2-\frac{9g_2^2}{4}-\frac{9g_1^2}{4}\right),\no\\
16\pi^2\frac{dy_n}{dt}&=&y_n\left(\frac{3}{2}\left(y_n^2-y^2\right)+\mathcal{N}\left(y^2+y^{c2}+y_n^2+y_n^{c2}\right)+3y_t^2-\frac{9g_2^2}{4}-\frac{9g_1^2}{20}\right),\no\\
16\pi^2\frac{dy_t}{dt}&=&y_t\left(\mathcal{N}\left(y^2+y^{c2}+y_n^2+y_n^{c2}\right)+\frac{9y_t^2}{2}-8g_3^2-\frac{9g_2^2}{4}-\frac{17 g_1^2}{20}\right),\no\\
16\pi^2\frac{d\lambda}{dt}&=&\lambda\left(24\lambda-9g_2^2-\frac{9g_1^2}{5}+12y_t^2+4\mathcal{N}\left(y_n^2+y_n^{c2}+y^2+y^{c2}\right)\right)
-2\mathcal{N}\left(y^4+y^{c4}+y_n^4+y_n^{c4}\right)-6y_t^4\no \\
&&+\frac{3}{8}\left(2g_2^4+\left(g_2^2+\frac{3g_1^2}{5}\right)^2\right). 
\eeq
The RGEs for $y^c$ and $y_n^c$ are similar to that for $y$ and $y_n$. 
The gauge beta functions are
\beq \label{eq:betagauge}
b_1=
\frac{41}{10}+\frac{6\mathcal{N}}{5},\;\;
b_2=
-\frac{19}{6}+\frac{2\mathcal{N}}{3},\;\;
b_3=
-7.\eeq
%

\paragraph{Vector doublets + triplet (``wino-higgsino").} For our ``wino-higgsino" scenario, including $\mathcal{N}$ copies with identical couplings and allowing for an additional singlet $n$, the relevant RGEs read~\cite{Giudice:2011cg}
\ba
16\pi^2\frac{dy}{dt}&=&y^cy_ny_n^c+y\left(\frac{5}{4}y^2+\frac{1}{4}y_n^2-\frac{1}{2}y^{c2}+\frac{\mathcal{N}}{2}\left(3y^2+3y^{c2}+y_n^2+y_n^{c2}\right)+3y_t^2-\frac{9}{20}g_1^2-\frac{33}{4}g_2^2\right),\no\\
16\pi^2\frac{dy_n}{dt}&=&3 y_n^c yy^c+y_n\left(\frac{3}{4}y_n^2+\frac{3}{2}y_n^{c2}+\frac{3}{4}y^2+\frac{\mathcal{N}}{2}\left(3y^2+3y^{c2}+y_n^2+y_n^{c2}\right)+3y_t^2-\frac{9}{20}g_1^2-\frac{9}{4}g_2^2\right),\no\\
16\pi^2\frac{dy_t}{dt}&=&y_t\left(\frac{\mathcal{N}}{2}\left(3y^2+3y^{c2}+y_n^2+y_n^{c2}\right)+\frac{9}{2}y_t^2-8g_3^2-\frac{17}{20}g_1^2-\frac{9}{4}g_2^2\right),\no\\
16\pi^2\frac{d\lambda}{dt}&=&\lambda\left(24\lambda+2\mathcal{N}\left(3y^2+3y^{c2}+y_n^2+y_n^{c2}\right)+12y_t^2\right)-\frac{\mathcal{N}}{2}\left(5y^4+5y^{c4}+y_n^4+y_n^{c4}\right) \no \\
&&-2\mathcal{N}y_ny_n^cyy^c-\mathcal{N}(y^2+y_n^{c2})(y^{c2}+y_n^2)-6y_t^4 \no \\
&&-9\lambda \left(\frac{g_1^2}{5}+g_2^2\right)+\frac{27}{200}g_1^4+\frac{9}{20}g_2^2g_1^2+\frac{9}{8}g_2^4.
\ea
\beq \label{eq:betagaugetrip}
b_1=
\frac{41}{10}+\frac{2\mathcal{N}}{5},\;\;
b_2=
-\frac{19}{6}+2\mathcal{N},\;\;
b_3=-7.\eeq

We take as initial conditions, at a scale $\mu=100$ GeV,
\beq
g_1&=&0.36\sqrt{5/3},\;g_2=0.65,\;
g_3=1.2,\;y_t=0.99,\;\lambda=\frac{m_h^2}{2v^2}=0.129.
\eeq
%


The vacuum stability cutoff scale $\Lambda_{UV}$ is determined by~\cite{Isidori:2001bm}
\beq\lambda\left(\Lambda_{UV}\right)=\frac{2\pi^2}{3\log\left(\frac{H}{\Lambda_{UV}}\right)}=-0.065\left(1-0.02\log_{10}\left(\frac{\Lambda_{UV}}{100~\rm GeV}\right)\right),\eeq
with the Hubble constant $H=70\,{\rm km/s/Mpc}=1.5\cdot10^{-42}$~GeV. 
We comment that for the problem under study, Landau poles of the Yukawa couplings appear at much higher scales, beyond the scale where the vacuum instability sets in, posing no additional constraint.\\

\section{Estimates of the collider constraints}
\label{sec:limits}
In this section, we present our estimates of the bounds on $\sigma \times \epsilon$, where $\epsilon$ includes acceptance, trigger and identification efficiencies, efficiencies of the kinematical cuts and the branching ratio to the relevant final state. We use a standard CLs technique to obtain the bounds and  we take the number of observed events and the predicted background with its error from the CMS mulitlepton search at 7 TeV performed with an integrated luminosity of 5 fb$^{-1}$~\cite{Chatrchyan:2012ye}. 
To get the excluded cross section, we first construct the likelihood as
\be
L(n | n_s+n_b)=\mathcal{P}(n | n_s+n_b)G(n_b | n_b^{obs}, \sigma_b)\, ,
\ee
where $\mathcal{P}$ is a Poisson distribution and we take a gaussian ansatz $G$ for the background. Then we compute the $\mathrm{CL}_s$,
\be
\mathrm{CL}_s =\frac{\mathrm{CL}_{s+b}}{\mathrm{CL}_b}=\frac{P(n \le n_d | n_s + n_b)}{P(n \le n_d | n_b)}\, ,
\ee
where $P$ is the probability obtained marginalizing the likelihood and $n_d$ is the number of events observed in the data. Requiring $\mathrm{CL}_s \le 0.05$ fixes $n_s$ to its $95 \%$ C.L. excluded value. In this way we obtain the cross section limits in Table \ref{tab:range1}.

\begin{table}[h]
\begin{center}
\begin{tabular}{|c|c|c|c|}
\hline
Selection & obs & background & $(\sigma \epsilon)_{\mathrm{excl}}$(fb) \\
\hline
4$l$, MET $<50$~GeV, $H_T< 200$~GeV, $Z$ & 33 & $37\pm 15$ & 5.6 \\
\hline
3$l$+1$\tau$, MET $<50$~GeV, $H_T< 200$~GeV, $Z$ & 20 & $17\pm 5.2$ &  3.4\\
\hline
\end{tabular}
\caption{Estimated bounds on $(\sigma \epsilon)_{\mathrm{excl}}$ of multi-lepton final states~\cite{Chatrchyan:2012ye} (see Sec.~\ref{sec:col}). $l$ refers to electrons or muons and $Z$ denotes two opposite sign leptons from a $Z$ decay.}
\label{tab:range1}
\end{center}
\end{table}
\end{appendix}

\bibliography{ref}
\bibliographystyle{jhep}
\end{document}